\documentclass[10pt]{iopart}
\usepackage[english]{babel}
\usepackage[utf8]{inputenc}
\usepackage{bm}
\usepackage{amsfonts}
\usepackage{amsthm}
\usepackage{iopams}
\usepackage{graphicx}
\usepackage{epsfig}
\usepackage{caption}
\usepackage{subfig}
\usepackage{braket}
\usepackage{verbatim}

\newcommand{\beq}{\begin{equation}}
\newcommand{\eeq}{\end{equation}}
\newcommand{\beqa}{\begin{eqnarray}}
\newcommand{\eeqa}{\end{eqnarray}}
\newcommand{\ba}{\begin{array}}
\newcommand{\ea}{\end{array}}
{\left\lbrace\begin{array}{@{}l@{}}}%
{\end{array}\right.}

\begin{document}
\title{Nonequilibrium Kinetics of One-Dimensional Bose Gases}

\author{F. Baldovin$^{1,2,3}$, A. Cappellaro$^{1}$, E. Orlandini$^{1,2,3}$, 
and L. Salasnich$^{1,2,4}$}
\address{$^{1}$Dipartimento di Fisica e Astronomia ``Galileo Galilei'', 
Universit\`a di Padova, Via Marzolo 8, 35122 Padova, Italy,\\
$^{2}$CNISM, Unit\`a di Ricerca di Padova, Via Marzolo 8, 35122 Padova, 
Italy, \\
$^{3}$INFN, Sezione di Padova, Via Marzolo 8, 35122 Padova, Italy, \\
$^{4}$INO-CNR, Sezione di Sesto Fiorentino, Via Nello Carrara, 
1 - 50019 Sesto Fiorentino, Italy}

\begin{abstract}
We study cold dilute gases made of bosonic atoms, 
showing that in the mean-field one-dimensional regime 
they support stable out-of-equilibrium 
states. Starting from the 3D Boltzmann-Vlasov equation 
with contact interaction, 
we derive an effective 1D Landau-Vlasov equation under the condition 
of a strong transverse harmonic confinement. 
We investigate the existence of out-of-equilibrium 
states, obtaining stability criteria similar 
to those of classical plasmas. 
\end{abstract}

\section{Introduction} 

Bose \cite{bose} and Fermi degeneracy \cite{fermi} were achieved 
some years ago in experiments with ultracold 
alkali-metal atoms, based on laser-cooling and 
magneto-optical trapping. 
These experiments have opened the way to the 
investigation and manipulation of novel states of atomic matter, like 
the Bose-Einstein condensate \cite{bose-rev} and the superfluid Fermi 
gas in the BCS-BEC crossover \cite{fermi-rev}. 
Simple but reliable 
theoretical tools for the study of these systems  
in the collisional regime are the hydrodynamic 
equations \cite{bose-rev,sala-flavio}. 
Nevertheless, to correctly 
reproduce dynamical properties of atomic gases in the mean-field collisionless 
regime \cite{pedri,rella}, or in the crossover from collisionless to 
collisional regime \cite{tosi}, one needs the Boltzmann-Vlasov 
equation \cite{landau1,pines,kada,zaremba}. 
Indeed, the Boltzmann-Vlasov 
equation is believed to be the correct equation to investigate the kinetics 
of a generic quantum gas made of out-of-condensate atoms \cite{griffin}. 
For dilute and cold atomic gases the mean-field potential 
of the Boltzmann-Vlasov equation 
\cite{landau1,pines,kada} is proportional to the s-wave scattering 
length of the inter-atomic interaction and to the local density 
of the gas \cite{pedri,griffin}. In the mean-field collisionless regime, 
where the interaction time between atoms is much larger than 
the characteristic period of the analyzed phenomenon, one can safely 
neglect the collisional integral of the Boltzmann-Vlasov equation 
getting the Landau-Vlasov equation, also called Hartree-Vlasov 
equation \cite{landau1}. 

The collisionless regime is strongly 
enhanced for a bosonic gas in a quasi one-dimensional (1D) 
configuration. A 3D system is said quasi-1D when the single-particle 
axial energy is much smaller than the energy of the transverse confinement. 
 Quasi-1D systems are nowadays routinely engineered with ultracold atoms 
in optical potentials, with harmonic transverse confinement energies
much larger than the gas temperature or chemical potential \cite{langen}. 
Remarkably, in a strictly 1D system of identical particles 
no thermalization may possibly occur \cite{guery}; 
such a behavior has been observed with cold bosonic atoms 
trapped in a 1D optical lattice \cite{weiss}. 
Indeed, particles completely exchange
their energy in 1D binary elastic collisions, so, 
if they are indistinguishable, no sensible
outcome is produced by such collisions.  
For 1D bosonic gases the collisionless 
Vlasov equation can be safely used under the condition 
$\lambda_{dB} \ll d \ll l_c$ \cite{guery}, where $\lambda_{dB}=
(2\pi\hbar)/\sqrt{2\pi m  k_BT}$ 
is the de Broglie wavelength with $T$ the absolute temperature, 
$d=1/\rho$ is the average distance between particles with $\rho$ the 1D 
local density, and $l_c=\hbar/\sqrt{m g  \rho}$ is the correlation 
length with $g\rho$ the 1D mean-field potential and $g$ the 1D 
effective interaction strength of the inter-atomic potential. 
It is well know that in a 1D bosonic system Bose-Einstein 
condensation is forbidden but quasi-condensation is 
possible \cite{bose-rev,landau1}; these inequalities 
ensure that the 1D bosonic gas does not contain 
a quasi-condensate (the classical description is applicable) 
and that it does not acquire the fermionic properties characteristic
of the Tonks-Girardeau  gas \cite{guery}. 

In this paper we consider the consequences of a strong confinement 
in the transverse plane and, assuming $\lambda_{dB} \ll d \ll l_c$, 
we derive an effective 1D Landau-Vlasov 
equation from its 3D analog. The resulting 1D equation is similar 
to the 3D starting one, except for the renormalization 
of the inter-atomic coupling constant by the square of the 
characteristic length of the transverse confinement. 
Stimulated by the recent activity on momentum-space engineering 
for gaseous Bose-Einstein condensates \cite{engi}, we then 
apply methods from plasma physics to
prove the existence and stability of out-of-equilibrium states 
for cold bosonic atoms. Specifically, we analyze small fluctuations around 
stationary solutions of the 1D Landau-Vlasov equation. 
For fermions, it is well known that  zero-sound oscillations
describe the perturbations of the Landau-Vlasov equation 
around the equilibrium Fermi-Dirac distribution 
\cite{landau1}. Following the analysis developed by 
Penrose \cite{penrose} and Gardner \cite{gardner} 
in plasma physics,  
our investigation generalizes to out-of-equilibrium 1D bosons 
the original zero-sound Landau result.
We also give the dispersion relation for modes associated to broad classes
of nonequilibrium solutions of the Landau-Vlasov equation. 
Finally, we generalize the Penrose criterion of  
instability to double-peaked distributions of cold bosonic atoms.
Although standard in plasma physics, our results 
points out that cold one-dimensional bosonic 
gases offer another experimental playground in which 
the above phenomenology may be investigated.

\section{Landau-Vlasov equation and its dimensional reduction} 

In the absence of Bose-Einstein condensation, cold and dilute 
gases made of bosonic alkali-metal atoms of mass $m$ are well 
described by the Boltzmann-Vlasov equation \cite{landau1,pines,kada,zaremba} 
\beq 
\left[ {\partial \over \partial t} + {{\bf p}\over m} 
\cdot \nabla_{\mathbf{r}} 
- \nabla_{\mathbf{r}}\big( U_{ext}+U_{mf} \big) \cdot \nabla_{\bf p} \right] 
F = I_{coll}[F] \; ,   
\label{figo}
\eeq
where $F({\bf r},{\bf p},t)$ is the  single-particle phase-space 
distribution. Here, 
\beq 
U_{ext}({\bf r}) = V(x) + W({\bf r}_{\bot}) 
\eeq
is the external trapping potential, due to a generic potential $V(x)$ 
in the axial direction $x$ and to a harmonic potential 
\beq 
W({\bf r}_{\bot}) = {m \omega_{\bot}^2\over 2} (y^2+z^2) 
\eeq
with frequency $\omega_{\bot}$ in the transverse direction 
${\bf r}_{\bot}=(y,z)$; 
\beq 
U_{mf}({\bf r},t) = \gamma \ n({\bf r},t) 
\eeq 
is the mean-field potential due to the inter-atomic interaction, with 
\beq 
\gamma={8\pi\hbar^2\over m} a_s  
\eeq
the interaction strength and $a_s$ the s-wave scattering 
length of the interaction between dilute bosonic atoms \cite{bose-rev,guery}.  
Nonlinerarities in the left-hand side of Eq.~(\ref{figo}) arise from the 
3D local density $n({\bf r},t)$, which is obtained from the  
phase-space distribution $F({\bf r},{\bf p},t)$ as 
\beq 
n({\bf r},t) = \int d^3{\bf p}  
\ F({\bf r},{\bf p},t) \; ;  
\label{loc_den}
\eeq
a further spatial integration gives the total number $N$ of atoms:
\beq 
N = \int d^3{\bf r} \ n({\bf r},t) \; . 
\eeq 
The mean-field potential $U_{mf}({\bf r},t)$, which is 
linear in $a_s$, affects the 
streaming part of the Boltzmann kinetic equation, while the 
collision integral $I_{coll}[F]$, which is quadratic 
in the scattering length $a_s$, describes dissipative 
processes \cite{pedri}. The collisional integral can be treated 
within a general formulation \cite{gupta} or, more simply,  
within the relaxation-time approximation \cite{pedri,kerson,ruffo} 
\beq 
I_{coll}[F] \simeq - {1\over \tau} \left[ 
F({\bf r},{\bf p},t)-F_{eq}({\bf r},{\bf p}) \right] \; , 
\eeq
where $\tau$ is the relaxation time related to the average 
time between collisions, and $F_{eq}({\bf r},{\bf p})$ is the 
global equilibrium distribution. 
By definition, in the mean-field collisionless regime 
the collisional integral can be neglected: in this case, 
the phenomenon under investigation 
has a characteristic time much smaller than the 
relaxation time \cite{landau1}. 

We shall work in this collisionless regime and 
the 3D Boltzmann-Vlasov equation (\ref{figo}) becomes the so-called 
3D Landau-Vlasov (or Hartree-Vlasov) equation \cite{vlasov,landau-damp}:
\beq 
\left[ {\partial \over \partial t} 
+ {{\bf p}\over m} \cdot \nabla_{\mathbf{r}} 
- \nabla_{\mathbf{r}} \big( 
V(x) + W({\bf r}_{\bot}) + \gamma \ n({\bf r},t)  \big) 
\cdot \nabla_{\bf p} \right] F = 0 \; .  
\label{figo1}
\eeq
A further simplification is achieved assuming that the atomic 
sample is under a very strong transverse confinement due to a large 
frequency $\omega_{\bot}$ of the trapping harmonic 
potential~\cite{bose-rev,langen}. 
Consistently, we suppose that the transverse energy 
of the confinement is much larger than the average axial kinetic energy 
of atoms. Namely\footnote{To simplify our notations below, at variance 
with customary practice $p$ represents instead the $x$-component 
of the vector $\bf p$.}, 
\beq 
\hbar \omega_{\bot} \gg {\langle p^2\rangle \over 2m} \; . 
\label{aiutone}
\eeq
In such a way the 3D system is constrained to occupy the 
transverse ground state, which is described by a Gaussian 
probability density of spatial width $a_\bot=\sqrt{\hbar/(m\omega_{\bot})}$,
thus becoming practically one-dimensional (quasi-1D) \cite{rella,io-mazza}. 
We set 
\beq 
F({\bf r},{\bf p},t) = f(x,p,t) \ f_{\bot}({\bf r}_{\bot},{\bf p}_{\bot}) \; , 
\label{ansatz}
\eeq
where $f(x,p,t)$ is the time-dependent axial distribution function 
and $f_{\bot}({\bf r}_{\bot},{\bf p}_{\bot})= f_{\bot}(y,z,p_y,p_z)$ 
is the transverse distribution function, defined as  
\beq 
f_{\bot}({\bf r}_{\bot},{\bf p}_{\bot}) 
= {1\over \pi^2 \hbar^2} \ 
\exp{\left( - {{p_{\bot}^2\over 2 m} + W_{\bot}({\bf r}_{\bot}) 
\over {\hbar \omega_{\bot}\over 2}} \right)} \; 
\label{boh1}
\eeq
and such that 
\beq 
n_{\bot}({\bf r}_{\bot}) = \int d^2{\bf p}_{\bot} 
f_{\bot}({\bf r}_{\bot},{\bf p}_{\bot}) = {1\over \pi a_{\bot}^2} 
\exp{\left(-{(y^2+z^2)\over a_{\bot}^2}\right)} \; . 
\label{boh2}
\eeq
Inserting Eq. (\ref{ansatz}) into Eq. (\ref{figo1}) and taking into
account Eqs. (\ref{boh1}) and (\ref{boh2}), after integration over
$y$, $p_{y}$, $z$, $p_{z}$ we are left with the 1D Landau-Vlasov
equation
\beq 
\left( {\partial \over \partial t} + 
{p\over m} {\partial \over \partial x} 
- {\partial \over \partial x}\big( V(x) + g \rho(x,t) \big)   
{\partial \over \partial p} \right) f = 0 \; , 
\label{1d-hve-io} 
\eeq
where 
\beq 
\rho(x,t) = \int dp \ f(x,p,t) 
\label{density}
\eeq
is the axial local density and 
\beq 
g = {\gamma} \ \int d^2{\bf r}_{\bot} \ n_{\bot}({\bf r}_{\bot})^2  = 
\frac{\gamma}{2\pi a_{\bot}^2} = {4\hbar^2 a_s \over m \, a_{\bot}^2}
\label{gggno}
\eeq
is the renormalized 1D interaction strength. Notice also that $f(x,p,t)$ 
is normalized to the total number $N$ of atoms, 
namely 
\beq 
N = \int dx \, dp \ f(x,p,t) \; . 
\label{norma}
\eeq  
In summary, Eq. (\ref{1d-hve-io}) 
is fully reliable for a 1D bosonic gas of cold atoms under two conditions: 
\beq 
{4\pi\hbar^2 \rho^2 \over \langle p^2\rangle } \ll 1 \; , 
\label{fuc1}
\eeq
ensuring that the 1D Bose gas is not in the quasi-condensate 
regime, and 
\beq 
{m g \over \hbar^2 \rho} = {4 a_s\over \rho \, a_{\bot}^2} \ll 1 \; ,  
\label{fuc2}
\eeq
which implies that the 1D Bose gas is not in the Tonks-Girardeau 
regime \cite{guery}. 
We remind that a more general derivation of the 1D kinetic 
equation, valid for a Bose gas characterized by both a 
quasi-condensate and a thermal component \cite{zaremba}, 
is provided in Ref.~\cite{arahata}. 

A remark about the possibility of fulfilling conditions
(\ref{aiutone}), (\ref{fuc1}),  (\ref{fuc2}) in experiments like those reported in
Refs. \cite{bose-rev,langen,guery} is in order. 
One can consider an axially uniform gas of alkali-metal 
atoms with linear density $\rho = N/L$. 
Typical experimental numbers are 
$N\simeq 10^2$,
$a_s/a_{\bot} \simeq 10^{-4}$, $L/a_{\bot} \simeq 10^4$, 
so that Eq. (\ref{fuc2}) is satisfied with   
${4 a_s/(\rho \, a_{\bot}^2)} 
\simeq 10^{-2}$. 
With these experimental values, we have 
$2\pi N^2 \left({a_{\bot}/L}\right)^2 
\simeq 10^{-2}$. 
Thus, even conditions (\ref{aiutone}) and (\ref{fuc1}), which can be 
rewritten as 
\beq 
2\pi N^2 \left({a_{\bot}\over L}\right)^2  \ll 
{ {\langle p^2 \rangle \over 2m} \over \hbar \omega_{\bot}} \ll 1 \; , 
\eeq
are fulfilled with the choice 
\begin{equation}
\langle p^2\rangle/(2m)\simeq 10^{-1}\hbar \omega_{\bot} \; ,
\label{pt}
\end{equation} 
a reasonable request in experimental setups.
For the important parameters $N$ and $\frac{a_{s}}{a_{\perp}}$ 
Eqs. (\ref{aiutone}), (\ref{fuc1}),  (\ref{fuc2}) set the ranges 
$10 \le N \le 5\cdot 10^{2}$ and $10^{-4} \le \frac{a_{s}}{a_{\bot}} \le 5\cdot 10^{-2}$.

\section{Stationary states} 

In this section we further simplify our analysis by dropping  
the axial confinement, $V(x)= 0$. 
The absence of an external axial 
potential is clearly a simplification. However, it is possible 
to experimentally produce 1D configurations of ultracold atoms 
without axial confinement by using a 1D ring geometry with a large 
radius \cite{weiss} or, alternatively, by adopting a 
straight axial 1D configuration with two high barriers 
at the edges \cite{hulet}. 
Consequently, the 1D Landau-Vlasov equation becomes 
\beq 
\left( {\partial \over \partial t} + 
{p\over m} {\partial \over \partial x} 
- g \int d{\tilde p} {\partial f(x,{\tilde p},t)\over \partial x} \ 
{\partial \over \partial p} \right) f(x,p,t) = 0 \; .  
\label{1d-hve} 
\eeq
Formally, any stationary and spatially uniform phase-space 
distribution $f_0(p)$ satisfies Eq. (\ref{1d-hve}), 
but of course the mere existence of a solution does not implies its stability. 
In general it is not required for $f_0(p)$ to coincide with the
  thermal equilibrium distribution;  
  the subject of our investigations is instead the dynamical stability of
  experimentally-engineered (out-of-equilibrium) $f_0(p)$'s.

Let $f_0(p)$ be a general spatially uniform 1D distribution 
function; we can set 
\beq 
f(x,p,t) = f_0(p) + \delta f(x,p,t) \; ,  
\eeq
where $\delta f$ is considered a small perturbation around 
the stationary solution. 
At linear order, $\delta f$ satisfies
\beq 
\left( {\partial \over \partial t} 
+ {p\over m} {\partial \over \partial x} \right) \delta f(x,p,t) 
- g {\partial f_0(p)\over \partial p} \int d{\tilde p}\;{\partial\, {\delta f}
(x,{\tilde p},t)\over \partial x} = 0 \; .  
\eeq
Following Landau's steps in plasma theory, the stability of $f_0$ can be
assessed by considering the initial-value problem
\cite{sturrock:plasmatheory,nicholson:plasmatheory}
\begin{equation}
\label{eq: cauchy initial value version of linearized vlasov equation}
\frac{\partial \delta f}{\partial t} + \frac{p}{m}\frac{\partial \delta f}
{\partial x} - g\int d\tilde{p}\frac{\partial \delta f}{\partial x}
(x,\tilde{p},t) \frac{\partial f_0}{\partial p}(p) = \delta f_{0}\;\delta(t) \;,
\end{equation}
with the prescription $\delta f(x,p,t) = 0$ for $t<0$. 
Indeed, the Dirac $\delta$-function in time 
brings 
the system to the perturbed initial state $f=f_0+\delta f$, characterized by 
$\delta f(x,p,0) = \delta f_{0}(x,p)$.
The time evolution of the perturbation determines whether 
$f_0$ is stable ($\delta f$ damps away), unstable ($\delta f$ grows up), 
or marginally stable (the amplitude of $\delta f$ remains constant).

\begin{figure}
\centering
\includegraphics[scale=0.60]{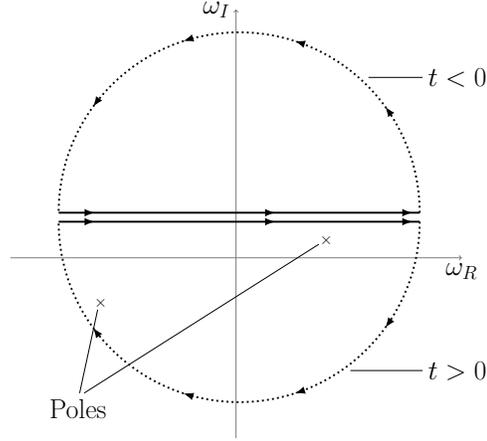}
\caption{Integration path $\Gamma_{\omega}$ for $t\lessgtr 0$.}
\label{figura1}
\end{figure}

This initial-value problem is most easily addressed through a
generalized Fourier-transform pair, involving a complex frequency
$\omega=\omega_R+\mathrm{i}\omega_I$:
\beq 
\delta\tilde f(k,p,\omega) = \int dx\,dt\; \delta{f}(x, p, t) 
\, \mathrm{e}^{-\mathrm{i}(kx - \omega t)};\,
\label{four}
\eeq 
\beq 
\delta f(x,p,t) = \int_{\Gamma_\omega} dk\,d\omega\; \delta\tilde{f}(k, p, \omega) 
\, \mathrm{e}^{\mathrm{i}(kx - \omega t)}\;.
\label{inv_four}
\eeq 
In order to satisfy the causality requirement ($\delta f=0$ for
$t<0$), the integration path $\Gamma_\omega$
in the complex $\omega$-plane can be chosen as a straight line
parallel to the real axis, lying above all singularities of
$\delta\tilde f$. At $\omega_R=\pm\infty$, such a path can be closed
by the portion of a circular line of infinite radius
in the upper (lower) $\omega$-plane for $t<0$ ($t>0$) 
(see Fig. \ref{figura1}).
By virtue of the residues theorem 
one thus sees that, among all possible modes, those that determine the
time evolution of the perturbation are related to the poles of
$\delta\tilde f$ in the complex $\omega$-plane. 
 
In view of Eq.~(\ref{eq: cauchy initial value version of 
linearized vlasov equation}), 
$\delta\tilde f$ satisfies
\begin{equation}
\label{eq: cauchy initial}
\bigg(-\omega + \frac{p}{m}k\bigg) \delta\tilde{f} - gk \int d\tilde{p}\;
\delta \tilde{f}(k,\tilde{p},\omega) \frac{\partial f_0(p)}{\partial p} 
= \frac{\delta \tilde{f}_0}{2\pi i}
\end{equation}
or
\begin{equation}
\label{eq: formal solution for cauchy initial value problem of lin. 
vlasov equation}
\delta \tilde{f}(k,p,\omega) = \frac{1}{2\pi i k} \frac{\delta \tilde{f}_0(k,p)}
{\big(\frac{p}{m} - \frac{\omega}{k}\big)} + \frac{g\frac{\partial f_0(p)}
{\partial p}}{\big(\frac{p}{m} -\frac{\omega}{k}\big)} \int d\tilde{p}\; 
\delta \tilde{f}(k,\tilde{p},\omega) \;.
\end{equation}
Integrating over $p$ we get
\begin{equation}
\label{eq: integral on momentum for the pertubation, cauchy problem}
\int dp \;\delta \tilde{f}(k,p,\omega) = -\frac{i}{2\pi k\,\epsilon(k,\omega)}
\int dp\; \delta\tilde{f}_0\bigg(\frac{p}{m} - \frac{\omega}{k}\bigg)^{-1} \;,
\end{equation}
with
\begin{equation}
\label{def: dielectric function}
\epsilon(k,\omega) = 1 - g \int dp\, \frac{\partial f_0(p)}{\partial p} 
\bigg(\frac{p}{m} - \frac{\omega}{k}\bigg)^{-1} \;.
\end{equation}
For non-pathological choices of the initial perturbation $\delta f_0$, the
integral at the r.h.s. of  
Eq.~(\ref{eq: integral on momentum for the pertubation, cauchy problem})
is well defined by appropriate choices of the integration path (see
below).  
According to 
Eq.~(\ref{eq: integral on momentum for the pertubation, cauchy problem}),
the singular behavior of $\delta\tilde f$ is then characterized by the
dispersion relation 
\begin{equation}
\epsilon(k,\omega)=0\;,
\label{nice}
\end{equation}
which identifies the poles of $\delta\tilde f$. 

\begin{figure}
\centering
\subfloat[][]
	{\includegraphics[scale=0.45]{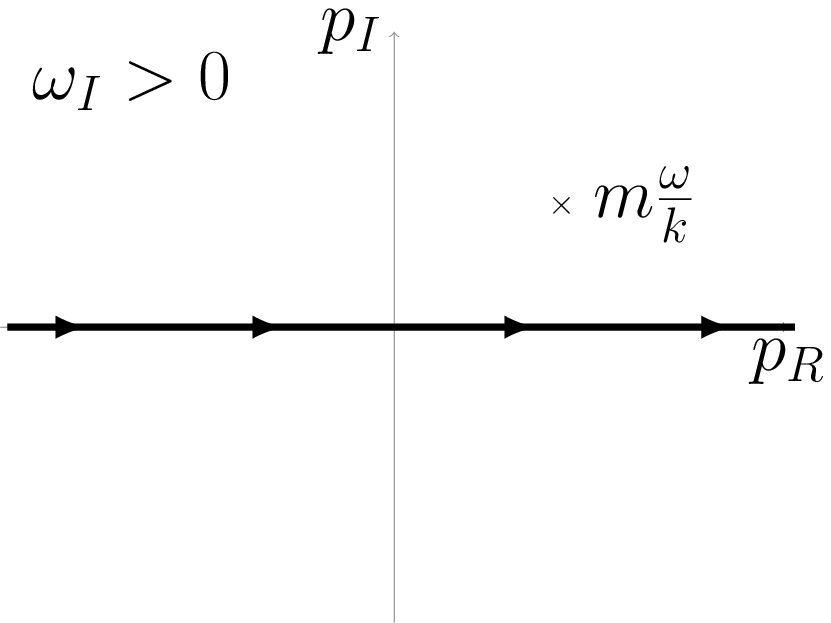}} \quad
\subfloat[][]
	{\includegraphics[scale=0.45]{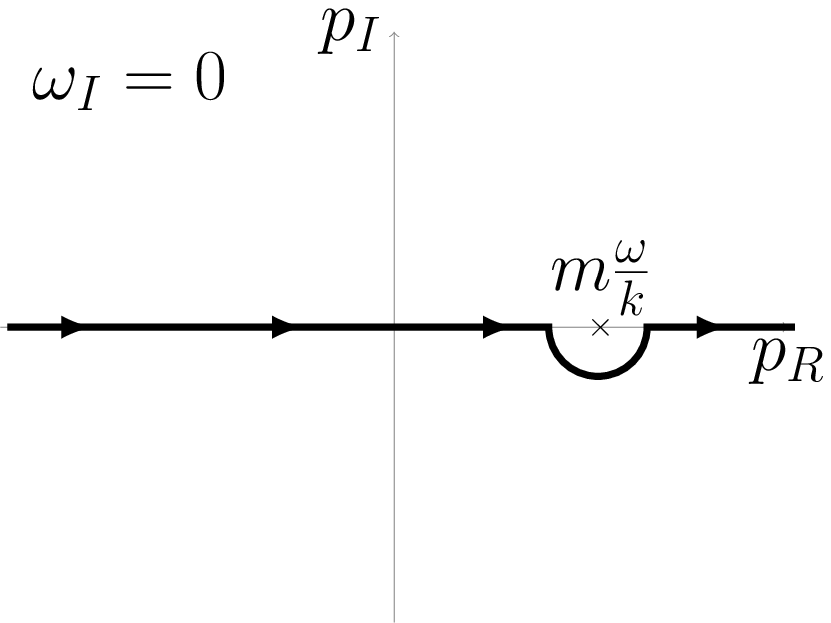}} \quad
\subfloat[][]
	{\includegraphics[scale=0.45]{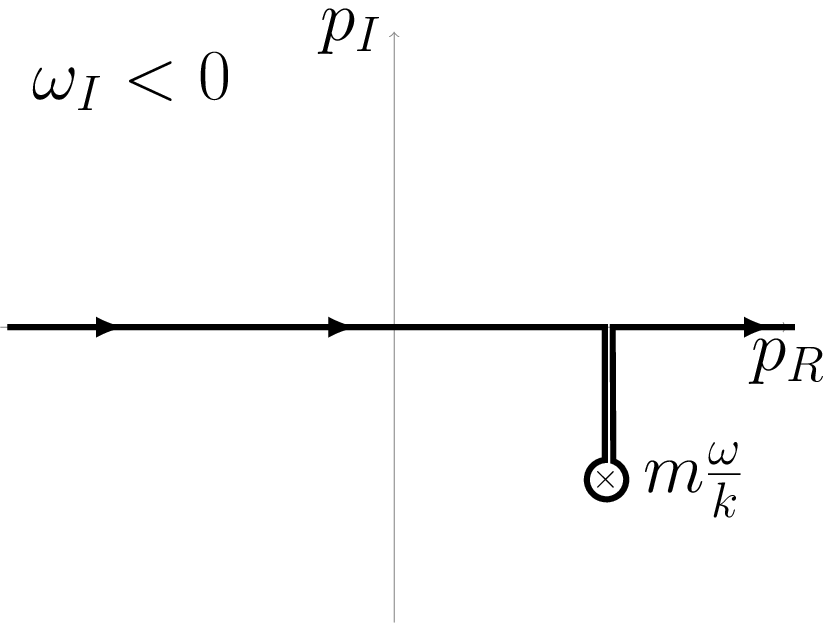}}
\caption{Integration paths in the complex $p$-plane for 
the analytical extension of $\epsilon(k,\omega)$.}
\label{figura 2}
\end{figure}

In analogy with plasma physics, $\epsilon(k,\omega)$ can be viewed as a 
{\it dielectric function}. In many-body physics it is called susceptibility 
or response function. 
Since the integration path $\Gamma_\omega$ is originally defined as a
straight line lying above all singularities, the dielectric function is
well defined for sufficiently large positive $\omega_I$.
In such a case, the integral at the r.h.s. of 
Eq.~(\ref{def: dielectric function}) lies on the real $p$-axis, below
the pole at $m\omega/k$. 
In order to close the integration path $\Gamma_\omega$ 
in the lower $\omega$-plane, to
evaluate the time evolution of the perturbation for $t>0$, we must
correspondingly extend analytically $\epsilon$ from the upper to the
lower $\omega$-plane. 
Complex analysis guarantees that 
the analytical extension is achieved by deforming the integration 
path on the real $p$-axis in such a way that it encircles the singularity at
$m\omega/k$ from below (see Fig. \ref{figura 2}).
If $\omega_I=0$ such a detour contributes for half the residue at the
pole; whereas if $\omega_I<0$ the contribution amounts to the full
residue. 

In summary, depending on the sign of $\omega_I$ we get three different
explicit dispersion relation $\epsilon(k,\omega)=0$ \cite{penrose,gardner}:
\beqa
1 &-& g \int_{-\infty}^{+\infty} dp\; {{f_0'}(p)\over p/m - \omega/k }  = 0 \quad 
\mbox{for} \; \omega_I>0 \; , 
\label{nice2}\\
1 &-& g\;\mathcal{P}\int_{-\infty}^{+\infty} 
dp\;{{f_0'}(p)\over p/m - \omega/k }  
 - \pi i \, g \, f_0'(m\omega/k) = 0 \quad 
\mbox{for} \; \omega_I=0 \; ,
\label{nice3}\\
1 &-& g \int_{-\infty}^{+\infty}dp\; {{f_0'}(p)\over p/m - \omega/k }  
-2\pi i \, g \, f_0'(m\omega/k) = 0 \quad 
\mbox{for} \; \omega_I<0 \; , 
\label{nice4}
\eeqa
where $f_0'(p)\equiv\partial f_0(p)/\partial p$, and 
the symbol $\mathcal{P}$ indicates the principal part of the integral.  
A solution of the dispersion relation with 
$\omega_I>0$ ($\omega_I<0$) corresponds to an exponentially growing 
(damping) mode. A solution with $\omega_I=0$ defines a
mode of constant amplitude. As a consequence, the stability of $f_0$ is
determined by the sign of the imaginary part of the solution of the
dispersion relation with the
largest $\omega_I$.

We remark that these relations are similar 
to the ones derived for plasmas from the 
Vlasov-Poisson equation \cite{vlasov,landau-damp}: in that case  
an additional factor $k^{-1}$ appears in the integrand. In our case, 
the equation $\epsilon(k,\omega) = 0$ shows instead that the dispersion 
relation between $\omega$ and $k$ is linear and gives implicitly 
the (complex) phase velocity of the mode,
\beq
c=\frac{\omega}{k},
\eeq 
as a function of the stationary 
solution $f_0(p)$ and of the parameters $m$ and $g$. 
It is also worth to point out that if one aims at discussing cases of 
singular distributions $f_0$ containing Dirac-deltas or step 
functions as in some of the below examples, results can be 
straightforwardly obtained without employing complex analysis.

\subsection{Zero-sound for 1D bosons} 

As a simple application, let us consider the stationary 
Fermi-Dirac-like distribution 
\beq 
f_0(p) = 
\Theta\left({p_F^2\over 2m} - {p^2\over 2m}\right) \; , 
\label{fermi}
\eeq
where $\Theta(x)$ is the step function 
and $p_F=\pi\hbar N/L$ is the 1D Fermi-like linear momentum fixed by the 
normalization (\ref{norma}), with $L$ the length of the axial domain.  
Under the condition $p_F\neq mc$ assuring $f_0'(mc)=0$, 
it is straightforward to extract from Eqs. (\ref{nice2}), 
(\ref{nice3}), (\ref{nice4}) the same result for the phase velocity 
of the perturbations: 
\beq 
c = \pm v_F \, 
\sqrt{ 1 + {2 g \over v_F} } \; ,    
\label{landau}
\eeq
where $v_F=p_F/m$ is the Fermi-like velocity. 
This formula gives nothing else than 
the familiar zero-sound velocity of a 1D Fermi gas 
\cite{landau1,pines}. 
Hence, zero-sound measurements 
can also be performed
for 1D bosons in the mean-field collisionless regime by tuning, for example,   
a stationary Fermi-like distribution of their momenta (\ref{fermi}). 
The two signs ($\pm$) simply mean that the perturbation 
is characterized by two waves which propagate along the $x$ axis 
in opposite directions. 
Equation (\ref{landau}) furthermore tells that such bosonic
Fermi-Dirac-like distribution is stable only for $g\geq -v_F/2$. 
We have thus shown that engineering the initial distribution 
of atomic bosons one gets the typical collisionless 
dynamics of a Fermi gas around its equilibrium configuration. 

\subsection{Stability of single-peak distributions and Landau damping} 

For single-peak distributions $f_0(p)$
one can state the following theorem, closely related 
to the analogous one developed by Penrose \cite{penrose} 
and Gardner \cite{gardner} for the Poisson-Vlasov 
equations of plasma physics 
\cite{sturrock:plasmatheory,nicholson:plasmatheory}.  

\vskip 0.4cm

\noindent 
{\bf Theorem}. 
{\it If the spatially uniform initial condition 
$f_0(p)$ has a single maximum and $g>0$, then $f_0(p)$ 
is linearly stable}.

\begin{proof}
Let us assume that the contrary is true, namely 
$c = \omega/k=c_R + \mathrm{i}c_I$ with $c_I>0$. 
We can rewrite Eq. (\ref{nice2}) as 
\beq
1 - g \int_{-\infty}^{+\infty} dp\ { (p/m -c_R + i c_I) f_0'(p) 
\over ( p/m -c_R)^2+ c_I^2 } = 0 \; .  
\eeq
This complex equation can be splitted into two real equations: 
\beqa
D_R(c_R,c_I) \equiv 
1 - g \int dp\ { (p/m -c_R) f_0'(p) 
\over ( p/m -c_R)^2+ c_I^2 } = 0 \; , 
\\
D_I(c_R,c_I) \equiv 
- g \int dp \ { c_I\;f_0'(p) 
\over ( p/m -c_R)^2+ c_I^2 } = 0 \; . 
\eeqa
Since $D_R(c_R,c_I)=0$ and $D_I(c_R,c_I)=0$, one has also 
\beq
D_R(c_R,c_I) + {p_0/m - c_R \over c_I} D_I(c_R,c_I) = 0 \; ,  
\label{eq_sing_max}
\eeq
where we choose $p_0$ as the single maximum of the $f_0$, satisfying 
in particular $f_0'(p_0)=0$. 
Equation~(\ref{eq_sing_max}) casts into
\beq
1 + g \int dp\ { (p_0/m-p/m) f_0'(p) 
\over ( p/m -c_R)^2+ c_I^2 } = 0 \; .
\label{eq_sing_max_2}
\eeq
In view of the single maximum at $p=p_0$,
\beq
(p_0/m-p/m) f_0'(p) \geq 0  \quad \forall p\in (-\infty,+\infty) \; . 
\eeq
This means that the integrand in Eq.~(\ref{eq_sing_max_2}) 
is non-negative for all values of $p$. Consequently, for $g>0$ 
Eq.~(\ref{eq_sing_max_2}) cannot be satisfied. 
This proves that our original assumption $c_I>0$ ($\omega_I>0)$
cannot be correct and  that $f_0$ is at least marginally stable. 
\end{proof}

Another situation typical of plasma physics which can be generalized to cold 
bosonic gases is that of weakly damped waves (or weakly unstable modes), 
i.e cases with $|\omega_I| \ll \omega_R$
\cite{sturrock:plasmatheory}. 
>From a physical point of view, this amounts to situations in which the
amplitude of the perturbations varies little in a time period. 
Because of the assumption $|\omega_I| \ll |\omega_R|$, a Taylor expansion 
of the dielectric function around the real value $\omega_R$ leads to
\begin{eqnarray}
\epsilon(k,\omega) &=& \epsilon (k, \omega_R + i\omega_I) \nonumber\\
&\simeq& 
\epsilon(k,\omega_R) + i \omega_I \frac{\partial \epsilon}{\partial 
\omega_R}(k,\omega_R)\;
\label{eq: first expansion}\\
&=&\epsilon_R(k,\omega_R) + 
i\epsilon_I(k,\omega_R) +\omega_I\bigg[i\frac{\partial \epsilon_R}
{\partial \omega_R}(k,\omega_R) - \frac{\partial \epsilon_I}
{\partial \omega_R}(k,\omega_R)\bigg].
\nonumber
\end{eqnarray}
Since $\omega_I$ is small, we may still  
make use of the dispersion relation in the form of Eq. (\ref{nice3}), 
namely
\begin{equation*}
\epsilon(k,\omega) = 1 - g\mathcal{P}\int dp\; \frac{\partial f_0(p)}
{\partial p}\bigg(\frac{p}{m} - \frac{\omega}{k}\bigg)^{-1} - 
i\pi g\frac{\partial f_0}{\partial p}\bigg|_{p=\frac{m\omega}{k}} \;.
\end{equation*}
Comparing Eq.~(\ref{nice3}) and Eq. (\ref{eq: first expansion}) we get
\begin{eqnarray}
\label{eq: second hase for the weakly damped waves}
\epsilon(k,\omega)&\simeq&
1 - g\mathcal{P}\int dp\;\frac{\partial f_0}{\partial p}(p)
\bigg(\frac{p}{m} - \frac{\omega_R}{k} \bigg)^{-1} - i\pi g
\frac{\partial f_0}{\partial p}(p)\bigg|_{p=\frac{m\omega_R}{k}} \; +
\nonumber \\
&& +\; i\omega_I \frac{\partial}{\partial \omega_R}
\bigg[ -g\mathcal{P} \int dp\; \frac{\partial f_0}{\partial p}(p)
\bigg(\frac{p}{m} - \frac{\omega_R}{k} \bigg)^{-1} \bigg] + 
\nonumber \\
&& + \omega_I \frac{\partial}{\partial \omega_R}\bigg[\pi g 
\frac{\partial f_0}{\partial p}(p)\bigg|_{p=\frac{m\omega_R}{k}} \bigg] \;.
\end{eqnarray}
To lowest order in $\omega_I/\omega_R$, we can neglect the last term
of the sum in the real part of the dielectric function.  
The dispersion relation $\epsilon(k,\omega)=0$ gives then two separate
equations: for the real part we have 
\begin{equation}
\label{eq: how to get the real part of omega}
1 - g\mathcal{P}\int dp\, \frac{\partial f_0}{\partial p}(p) 
\bigg(\frac{p}{m} - \frac{\omega_R}{k}\bigg)^{-1} = 0 \;;
\end{equation}
and for the imaginary one
\begin{equation}
\label{eq: how to get the imaginary part  of omega}
\omega_I = -\frac{\pi \frac{\partial f_0}{\partial p}
\big(\frac{m\omega_R}{k}\big)}{\frac{\partial }{\partial \omega_R}
\bigg[\mathcal{P}\int dp\, \frac{\partial f_0}{\partial p}(p)
\big(\frac{p}{m} - \frac{\omega}{k}\big)^{-1}\bigg] \bigg|_{\omega_I = 0}} \;.
\end{equation}

A simple exemplification of the above theorem and an
explicit calculation of the damping rate may be given in terms of
the stationary Maxwell-Boltzmann distribution 
\beq 
f_0(p) = \rho \ {1\over \sqrt{\pi} \overline{p}} e^{-p^2/\overline{p}^2} \; ,  
\label{mb-dis}
\eeq
where $\rho=N/L$ is 
fixed by Eq. (\ref{norma}), with $L$ the length of the axial 
domain and $\overline{p}$ the characteristic width of the 
distribution. 
Under the condition $\overline{p}\ll \sqrt{mg\rho}$ it is not difficult 
to obtain, 
by Eq.(\ref{eq: how to get the real part of omega}), 
the real part of the velocity $c_R=\omega_R/k$ as 
\beq 
c_R = \pm \sqrt{{g\rho\over m}\left( 1 + {3\over 2} 
{\overline{p}^2\over m g \rho}\right)} \; , 
\eeq
while, in view of Eq.(\ref{eq: how to get the imaginary part  of omega}),
the imaginary part $c_I=\omega_I/k$ reads 
\beq 
c_I = - \sqrt{\pi} \, e^{-3/2} \, 
{m g^2\rho^2\over \overline{p}^3} \, e^{-mg\rho/\overline{p}^2} \; .  
\eeq
This is a meaningful result: the negative sign of $c_I$ corresponds 
to Landau damping, the typical non-collisional phenomenon of wave-particle 
interaction; in the present case the wave is not an electrostatic 
\cite{landau-damp}, but a matter one.

In the limit of vanishing $\overline{p}$, 
the distribution (\ref{mb-dis}) becomes a Dirac delta function
\beq 
f_0(p) = \rho \ \delta\left(p\right) \; , 
\label{delta}
\eeq
while the  velocity becomes real and given by 
\beq 
c = \pm \sqrt{g\rho\over m} \;. 
\eeq
Namely, the zero-sound velocity associated to a strongly localized peak.

\subsection{Two-stream instability} 

We eventually address another interesting 
situation in which a spatially uniform distribution may not be 
stable, namely 
a case with a double-peak $f_0(p)$.
Arguably, in the  
simplest case of this kind $f_0(p)$ is a linear combination of two Dirac delta,
\beq
f_0(p) = \frac{\rho}{2} \left [\delta(p-p_0) + \delta(p+p_0) \right ] \; , 
\label{two_deltas} 
\eeq
representing two streams of particles propagating with the same
velocity in opposite directions. 
By considering the dispersion relation, Eq. (\ref{nice}), for such
distribution with $c=\omega/k\ne p_0/m$, we obtain
\beq
c^2_{\pm} = \frac{1}{m^2} \left (p_0^2 +\frac{1}{2} g m \rho \pm 
\frac{1}{2}\sqrt{8 m g \rho p_0^2 + m^2 g^2\rho^2} \right) \; . 
\eeq
Each choice of the $\pm$ sign gives two roots for $c$. If $c^2 >0$
the two roots are real and, since $k \in \mathbb{R}$, this gives
$\omega = \omega_R = \pm k \sqrt{c^2}$. 
This is for instance the case of 
the solution $c^2_+ = \frac{1}{m^2} \left (p_0^2 +\frac{1}{2} m g \rho +
\frac{1}{2}\sqrt{8 m g \rho p_0^2 + m^2 g^2 \rho^2} \right)$ (with $g>0$).
Such a solution describes a circumstance in which 
$f_0$ is marginally stable, the perturbations
maintaining constant amplitude and propagating at a velocity 
\beq
c = \pm \sqrt{\frac{1}{m^2} 
\left (p_0^2 +\frac{1}{2} m g \rho + 
\frac{1}{2}\sqrt{8 m g \rho p_0^2 + m^2 g^2 \rho^2} \right)} \; .
\eeq
A perhaps more intriguing phenomenology is related to the 
solution $c^2_-$, which may turn negative
for some values of the parameters. 
With $c^2_- < 0$ the two roots for $c=\omega/k$ are purely 
imaginary, equal in magnitude, and opposite in sign. 
For each given $k$, the positive imaginary root thus characterizes
a perturbation exponentially growing in time at a rate $|c_{-}|$.  
The instability condition $c^2_{-}<0$ corresponds to the interval
\beq
-\sqrt{m g \rho} < p_0 < \sqrt{m g \rho} \; . 
\eeq
Under these premises, the maximum growth rate $|c_{ins}^{max}|$ 
is found by maximizing 
\begin{equation}
c^2_{-}(p_0) = \frac{1}{m^2} \left (p_0^2 +\frac{1}{2} m g \rho -
\frac{1}{2}\sqrt{8 m g \rho p_0^2 + m^2 g^2 \rho^2} \right) \;, 
\end{equation}
and is achieved for $p_0 = \frac{1}{4}\sqrt{6 m g \rho}$
with  
\beq
|c_{ins}^{max}| = \sqrt{\frac{g \rho}{8m}} \;. 
\eeq

More general two-stream instability conditions are given by 
the following theorem, generalizing the Penrose criterion 
of plasma physics 
\cite{penrose,sturrock:plasmatheory,nicholson:plasmatheory} to
cold bosonic gases.  

\vskip 0.4cm

\noindent 
{\bf Theorem}. 
{\it  If the spatially uniform initial condition 
$f_0(p)$ has two local maxima, a local minumum at $p=p_{min}$,  and $g>0$,
then $f_0(p)$ 
is unstable under the condition 
\begin{equation}
\mathcal{P}\int_{-\infty}^{+\infty} dp \frac{f_0(p) - f_0(p_{min})}
{(p - p_{min})^2} >\frac{1}{g \ m} \; .
\label{eq_penrose}
\end{equation}
}

\begin{proof}
The instability of $f_0$ corresponds to the existence of zeros of
$\epsilon(k,\omega)$  
with a positive imaginary part. 
If $\epsilon(k,\omega)$ has no poles in the upper $\omega$-plane, the
number of such zeros can be computed through the integral
\begin{equation}
\label{eq: number of zeros of dielectric function}
\frac{1}{2\pi i}\int_{\Gamma_{\omega}} d\omega\; \frac{1}
{\epsilon(k,\omega)}\frac{\partial \epsilon}{\partial \omega}(k,\omega) \; ,
\end{equation}
where the contour $\Gamma_{\omega}$ runs along the real $\omega$-axis 
and is then closed counterclockwise 
by a semicircle of infinite radius in the upper half-plane.  
The contour $\Gamma_\omega$ in the $\omega$-plane maps to a
contour $\Gamma_\epsilon$ in the $\epsilon$-plane obtained by 
evaluating $\epsilon(k,\omega)$ at every point of $\Gamma_{\omega}$.
The distribution $f_0$ is unstable if and only if 
$\Gamma_{\epsilon}$ encloses the origin.
In the present case of two-peak $f_0$ this is possible only if the
real part of $\epsilon$ evaluated in correspondence of  
$\omega=\omega_{min}\equiv k\,p_{min}/m$ is negative.
An integration by part then leads to the fact that
a necessary and sufficient 
condition for the instability of $f_0$ is given by
Eq.~(\ref{eq_penrose}). 
\end{proof}

\begin{figure}[ht]
\centerline{\epsfig{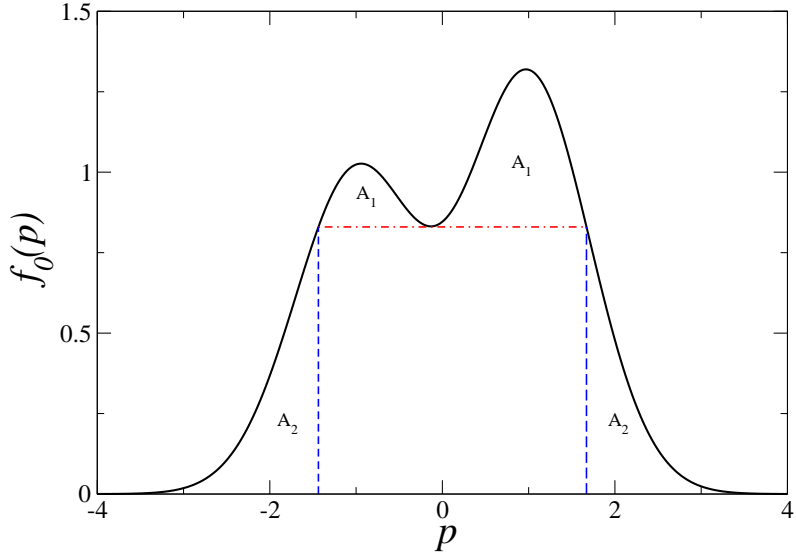}}
\caption{Initial distribution $f_0(p)$ with two peaks. 
Here the local minimum is at $p_{min}=0$. The configuration is unstable 
if $A_{1}/A_{2}>1/(g m)$, where $A_1$ is the area between the portion of  
$f_0(p)$ containing the two maxima and the horizontal dot-dashed line, 
and $A_2$ is the area of the ``external'' regions delimited by the rest of 
$f_0(p)$, the two vertical dashed lines, and the horizontal axis.}
\label{fig:twopeaks}
\end{figure}

The instability criterion in Eq.~(\ref{eq_penrose}) has an interesting
graphical visualization \cite{sturrock:plasmatheory}. 
Consider the plot $f_0(p)$ of Fig. \ref{fig:twopeaks} and an horizontal 
line passing through $f_0(p_{min})$. 
Further consider the vertical segments between the $p$-axis and the
intersection points of the horizontal line with $f_0(p)$.  
These vertical lines identify two ``external'' regions.
The left one is delimited by $f(p)$, the portion of $p$-axis including
$p=-\infty$, and the left vertical line.
Analogously, 
the right one is delimited by $f(p)$, the portion of $p$-axis including
$p=+\infty$, and the right vertical line.
The criterion in Eq.~(\ref{eq_penrose}) means that the area between
$f(p)$ and the horizontal line must outweight of at least a factor $1/gm$ the
area of the ``external'' regions. In other words, the dip at the
minimum of $f_0$ must be sufficiently deep.

At latter, it is pertinent to observe that distributions with two distinct 
momentum components are now frequently realized with ultracold atoms 
by means of two-photon Raman and Bragg transitions \cite{weid}. 
Indeed, by adjusting the 
angle and the frequency difference of the two laser beams driving 
the transition, each of the two peaks of the imparted momentum can be 
calibrated within the range 
$[-1, 1] \times 2 \hbar k_L$, where $\pm\hbar k_L$ are the
momenta of the laser photons. 

\section{Axial confinement}

Neglecting the axial potential is an assumption
that may hinder important dynamical effects (see, e.g.,
Ref. \cite{jackson} for a discussion within the context of plasmas).  
Since most of the experiments are carried out in the presence of 
an external potential $V(x)$, it is indeed interesting to 
study numerically the stability of nonequilibrium distributions for 
given profiles of $V(x)$. 

\begin{figure}[ht]
\centerline{\epsfig{file=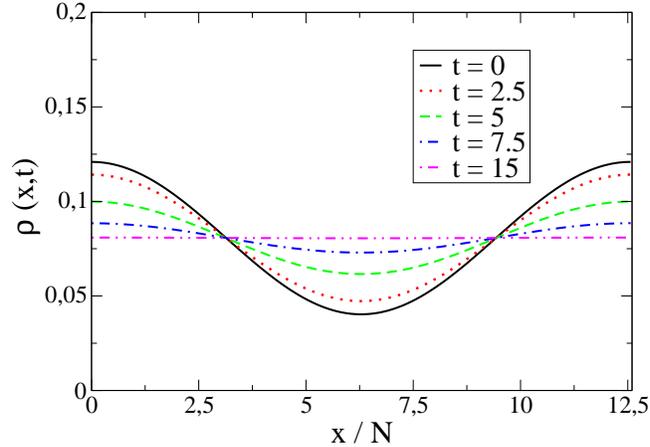,width=9.3cm,clip=}}
\caption{Vlasov time-evolution of the 1D bosonic density profile
  $\rho(x,t)$.  In the simulation units the initial condition is given by 
  Eqs.~(\ref{mb-dis}), (\ref{eq_numerical_test_for_damping}),
  with $\rho=0.084$, $N = 100$, $\overline{p}=0.71$, $\alpha=0.2$, $k=0.5$,
  $L=12.6\cdot 10^2$, and $p_{max} = 3$. The interaction
  strength is fixed at $g=0.1$.  
  The phase space $(x,p)$ has been discretized by a 2D grid of lattice
  spacing $\Delta x = 0.036$ and $\Delta p = 0.012$, whereas for the
  integration time we have chosen $\Delta t = 0.005$.  Curves
  correspond to increasing values of time $t$.  As time passes by, the
  initial perturbation is damped out and a stable uniform spatial
  density is reached.}
\label{fig:damping}
\end{figure}

To integrate numerically Eq. (\ref{1d-hve}) we adopt a semi-Lagrangian
scheme with cubic spline interpolation method~\cite{semilagrangian}, adapted
to the 1D Landau-Vlasov equation for cold bosonic atoms.
In order to check the validity of the numerical implementation as well as to 
compare the relevant time scales of the problem we first analyze the occurrence 
and time behavior of Landau damping in absence of an axial 
confinement ($V(x)=0$).
In particular we consider the time evolution of the 
initially perturbed stationary state $f_0$:
\begin{equation}
\label{eq_numerical_test_for_damping}
f(x,p,0) = f_0(p)[1 + \alpha\cos(kx)],\quad x\in [0,L], \;p\in 
[-p_{max},p_{max}].
\end{equation}
where $f_0$ is given by  Eq. (\ref{mb-dis}).

In order to compare the simulation results with experiments we apply
the following procedure.
Let us call, respectively, $\mathrm{u_L}$, $\mathrm{u_T}$,
$\mathrm{u_M}$ the length, time, and mass units employed in our
simulations. 
The relation between these units and the physical ones can be obtained
by equating the values of $\rho(x,t)$, $g$, $\overline{p}$ of the
simulations with their typical experimental counterparts, and solving
in $\mathrm{u_L}$, $\mathrm{u_T}$, $\mathrm{u_M}$ the resulting system
of equations.
If we consider reasonable experimental numbers as $N \simeq 100$,
$a_{\perp}\simeq 10^{4} a_{s}$ with $a_{s} \simeq 10^{-10}$ m, we get
$\rho(x,t) \simeq 10^{4}$ m$^{-1}$.
Furthermore, taking a Li atom of mass $m\simeq 9\cdot10^{-27}\;{\rm kg}$, we have
$g\simeq5\cdot10^{-41}\;{\rm kg\;m^3\;s^{-2}}$.
Finally, adopting Eq. (\ref{pt}),
$\overline{p}\simeq4\cdot10^{-29}\;{\rm kg\;m\;s^{-1}}$.
Correspondingly, we thus find 
$1\;\mathrm{u_L}= 8.4 \cdot 10^{-6}\;{\rm m}$, 
$1\;\mathrm{u_T}= 0.2\;{\rm s}$,
$1\;\mathrm{u_M}= 3\cdot10^{-27}\;{\rm kg}$.

In Figure \ref{fig:damping} we show the density profile $\rho(x,t)$ 
taken at different time steps.
As expected by the considerations given in section 3, the initial perturbation 
damps down to a spatially uniform density profile with a damping time
$\tau_D\simeq 10 \;\mathrm{u_T}=2\;\mathrm{s}$.
Nowadays, experimental duration time of the order of seconds are
indeed at reach. 

Although collisions are effectless in strictly 1D configurations,
they still provide a thermalization mechanism to quasi-1D
systems. It is then interesting to compare the typical collision time with the 
the Landau damping time $\tau_D$.
The collision time $\tau_{c}$ can be estimated as \cite{kerson}
\begin{equation}
\tau_{c} \simeq \frac{1}{n \sigma \overline{v}},
\end{equation}
where $\overline{v}=\overline{p}/m$ 
is the characteristic velocity of the particles,
and $\sigma=8\pi a_s^2$ is the cross-section. 
For our quasi-1D setup we may take  $n=\rho/(2\pi a_{\bot}^2)$ with $\rho=N/L$, 
and $\tau_c$ then reads $\tau_c=1/(4 \rho (a_s/a_{\bot})^2
\overline{p}/m)$. 
Adopting this time natural units with $\hbar=m=1$, from Eq. (\ref{gggno}) one finds 
$a_s^2=g^2a_{\bot}^4/16$ and consequently $\tau_c=4/(\rho g^2L^2 
\overline{p} (a_{\bot}/L)^2)$.
By using the values in the caption of Fig. \ref{fig:damping} we then
obtain $\tau_c \simeq 7\cdot 10^{3}\,(L/a_{\bot})^2$. 
We notice that the collisional time is proportional to the
$L/a_{\perp}$; in order to obtain realistic quasi-1D configurations,
this ratio has to be engineered such that $L/a_{\perp} \simeq 10$ or
more. If we impose that this ratio is $10$, then the collision time is
$\tau \simeq 10^{5}$, four orders of magnitude larger than the damping 
time $\tau_{D}$.

\begin{figure}[ht]
\centering
	{\epsfig{file=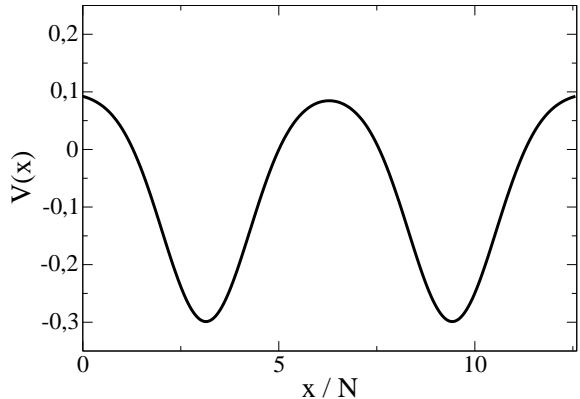,width=8.5cm,clip=}} 
	\caption{The axial potential defined in Eq.~(\ref{potential}) 
	for $s = 1.58$, $x_0=L/4$, $x_1 = 3L/4$, $N=100$, and system size
	$L=12.6\cdot 10^2$.}
\label{fig:potential}
\end{figure}

\begin{figure}[ht]
\centering
	{\epsfig{file=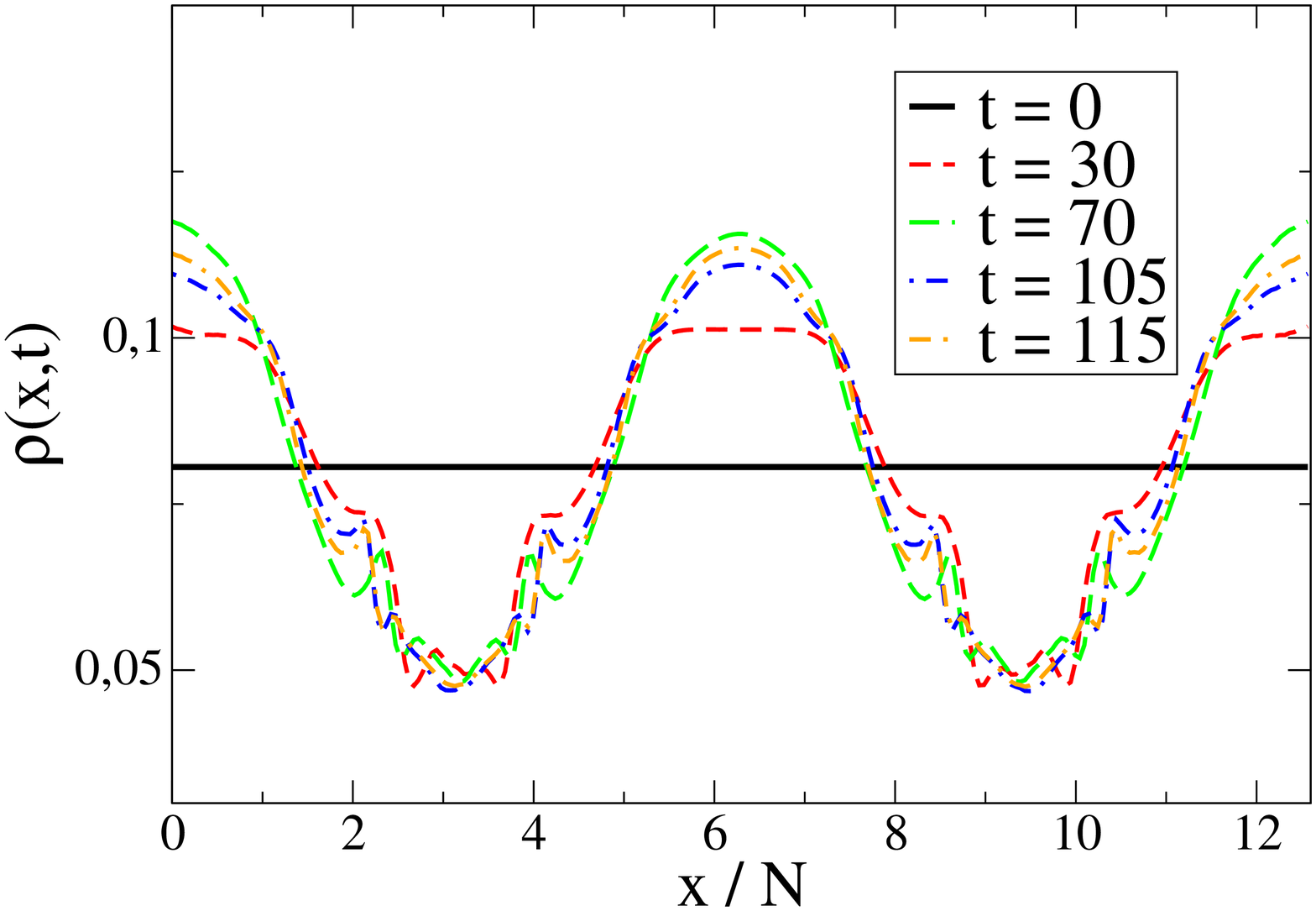,width=9.5cm,clip=}} \quad
	{\epsfig{file=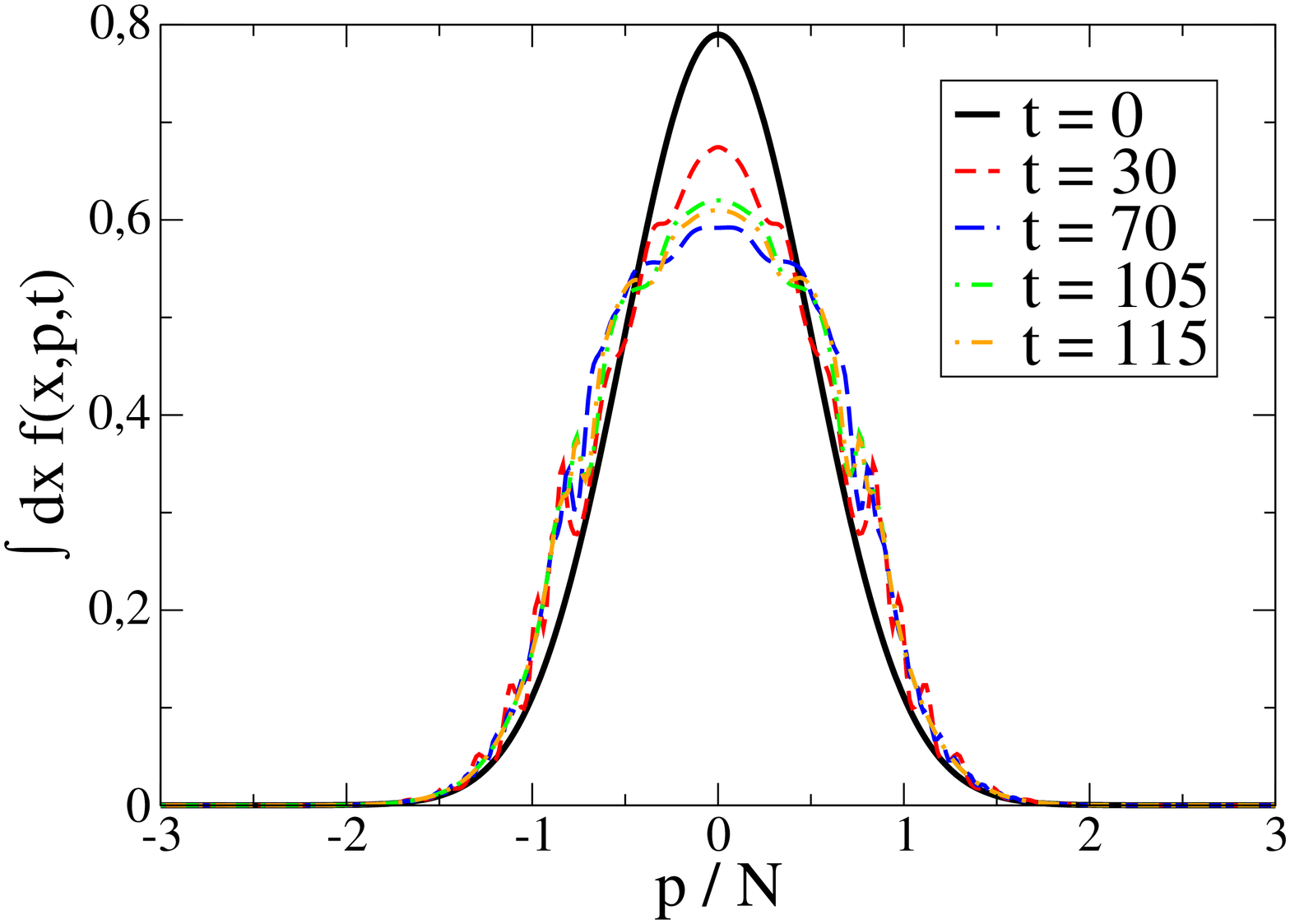,width=9.5cm,clip=}}
	\caption{Vlasov time-evolution in the presence of the external
	  potential $V(x)$ in Fig.~\ref{fig:potential}.  In the
	  simulation units, the initial 1D distribution (see solid and
	  thicker curves in both panels) is set to be uniform along
	  $x$ with density $\rho=0.084$ (top panel) and Gaussian in
	  the momenta space as in Eq. (\ref{mb-dis}) with
	  $\overline{p}=0.71$ and $p_{max} = 3$ (bottom panel).  As in
	  Figure \ref{fig:damping}, $N = 100$, $g=0.1$, $\Delta x =
	  0.036$, $\Delta p = 0.012$ and $\Delta t = 0.005$.  In both
	  panels the dashed curves refer to the density $\rho(x,t)$
	  and the reduced probability distribution $\int f(x,p,t) dx$
	  profiles taken at different times.  Note that, as $t$
	  increases, the curves tend to stabilize around a non-Gaussian
	  shape. }
\label{fig:density}
\end{figure}

We now consider the system in the presence of a symmetric external axial potential 
of double-well shape:
\begin{equation}
V(x) = -\frac{1}{\sqrt{2\pi}}\left[e^{-\frac{(x-x_0)^2}{s^2}} 
+ e^{-\frac{(x-x_1)^2}{s^2}}\right]
\label{potential}
\end{equation} 
(see Fig.~\ref{fig:potential}). 
The effect of this potential on an initially uniform 
spatial profile and momenta distribution as in Eq. (\ref{mb-dis}) is  reported
in Fig. \ref{fig:density}: it is readily seen that 
Landau damping provide a relaxation mechanism alternative to
collisions which drives the system to a stationary
state reproducing the double-well shape of the potential. 
This implies that the switch on of a potential acts as a
perturbation which interacts with the gas atoms in a sort of
wave-particle energy exchange, similarly to 
the standard interpretation of Landau damping.

\section{Conclusions} 

Within a mean-field collisionless regime, 
we have shown that stable out-of-equilibrium 
states can be produced with 1D Bose gases made of cold 
alkali-metal atoms. This prediction is based on the analysis 
of the effective 1D Landau-Vlasov equation derived from the 3D Boltzmann 
equation under the condition of a strong transverse confinement. 
By investigating the linearized 1D Landau-Vlasov equation, we have obtained  
theorems addressing the stability of such out-of-equilibrium solutions. 
Emphasizing analogies with plasma physics,
we have been able to point out the existence of a variegate
phenomenology also applicable to experimental design for cold 1D 
bosonic gases. Results include zero-sound transmission for 
bosonic species, Landau damping, and two-stream
instability. On longer time scales controlled by the length-to-width
ratio of the quasi-1D configuration, residual correlations and 
many-atom collisional effects are expected to erase 
such phenomena as a side effect of the approach to equilibrium. 
Finally, as a simple but meaningful example, we have shown that 
switching on suddenly an external double-well potential $V(x)$ 
the atomic cloud goes towards a new out-of-equilibrium 
stationary configuration. Similar configurations, obtained 
with different external potentials $V(x)$ and initial conditions, 
are presently under investigation. 

{\it Acknowledgments}. 
The authors thank David Guery-Odelin, Francesco Minardi, 
Roberto Onofrio, Flavio Toigo for 
useful suggestions and acknowledge Ministero Istruzione
Universita Ricerca (PRIN project 2010LLKJBX) for partial support.

\end{document}